# Adversarial Robustness Study of Convolutional Neural Network for Lumbar Disk Shape Reconstruction from MR images

Jiasong Chen[a], Linchen Qian[a], Timur Urakov[b], Weiyong Gu[c], Liang Liang*[a]

[a]Dept. of Computer Science, [b]Dept. of Neurological Surgery, and [c]Dept. of Mechanical & Aerospace Engineering at University of Miami, Coral Gables, FL 33146. *Email: liang@cs.miami.edu

**ABSTRACT**

Machine learning technologies using deep neural networks (DNNs), especially convolutional neural networks (CNNs), have made automated, accurate, and fast medical image analysis a reality for many applications, and some DNN-based medical image analysis systems have even been FDA-cleared. Despite the progress, challenges remain to build DNNs as reliable as human expert doctors. It is known that DNN classifiers may not be robust to noises: by adding a small amount of noise to an input image, a DNN classifier may make a wrong classification of the noisy image (i.e., in-distribution adversarial sample), whereas it makes the right classification of the clean image. Another issue is caused by out-of-distribution samples that are not similar to any sample in the training set. Given such a sample as input, the output of a DNN will become meaningless. In this study, we investigated the in-distribution (IND) and out-of-distribution (OOD) adversarial robustness of a representative CNN for lumbar disk shape reconstruction from spine MR images. To study the relationship between dataset size and robustness to IND adversarial attacks, we used a data augmentation method to create training sets with different levels of shape variations. We utilized the PGD-based algorithm for IND adversarial attacks and extended it for OOD adversarial attacks to generate OOD adversarial samples for model testing. The results show that IND adversarial training can improve the CNN robustness to IND adversarial attacks, and larger training datasets may lead to higher IND robustness. However, it is still a challenge to defend against OOD adversarial attacks.

**Keywords:** deep neural network, adversarial robustness, in-distribution, out-of-distribution, lumbar disk image

## 1. INTRODUCTION

Deep neural networks (DNNs), especially convolutional neural networks (CNNs), have become the method of choice for medical image analysis, automating the entire analysis process in an end-to-end fashion while being as accurate as human doctors in many applications [1-3]. Several DNN-based medical image analysis systems have been FDA-cleared [4] and will make impacts on human lives. Despite the progress, challenges remain to build DNNs as reliable as human expert doctors. From the perspective of technology robustness, two challenges need to be resolved:

The first challenge is DNN robustness to noises. It has been found that DNNs are not robust to a type of noise, called adversarial noise [5]. Adversarial noise was first discovered by Szegedy *et al.* 2014 [6] and then explained by Goodfellow *et al.* 2014 [7]. Adversarial noise can significantly affect robustness of DNNs for a wide range of image classification applications [5, 8], such as handwritten digits recognition [9], animal classification [10], human faces recognition [11], lung X-ray image classification [12], and even traffic sign detection [13]. Due to the severity of this robustness issue, defense methods were proposed to fight against adversarial attacks [5, 14, 15]. The most popular strategy is adversarial training [6, 7, 15, 16], which adds adversarial noise to images and uses the noisy images for DNN training to improve robustness. The current defense methods were mainly tested on "standard" image datasets, such as MNIST, CIFAR, and ImageNet. The impact of adversarial noise on CNN robustness for the application of spine MR image analysis, to our knowledge, has not been studied except in our work.

The second challenge is the detection of out-of-distribution (OOD) samples. If a sample does not follow the distribution of the training set (i.e., not similar to any sample in the training set), then the sample is OOD. If a sample follows the distribution of the training set, then the sample is in-distribution (IND), and all of the samples in the training set are IND. Given an OOD sample as input, the output of a DNN may become meaningless. For example, it will be meaningless to use a DNN classifier, which is trained on spine MR images, to make disease diagnosis from chest CT images or some random noise images. For a DNN-based medical system that is expected to work independently without human intervention, it must have the capability of detecting OOD samples: once an OOD sample is detected, it will notify the user and refuse to make a diagnostic decision. There have been several attempts to develop methods for OOD sample detection in classification applications, but our recent work shows that none of these methods are effective: they cannot distinguish between IND samples and OOD samples generated by OOD adversarial attacks [17]. For medical image analysis, reconstruction autoencoders were used for OOD detection with the assumption that the reconstruction error of an OOD sample will be much higher than the reconstruction errors of IND samples [18]. In this study, we will show that this reconstruction-based OOD detection method is ineffective for our application of spine image analysis.

We were interested in lumbar spine image analysis for disk degeneration assessment. Human intervertebral discs [19] undergo a process of profound degeneration as early as the age of 12, and this process may manifest as discogenic low back pain, disc herniation, spinal stenosis, and/or spondylolisthesis, which may require surgical or non-surgical treatments to reduce pain and restore normal functions. Using magnetic resonance (MR) imaging, disc degeneration can be revealed in terms of disk geometry deformation and signal strength degradation. Currently, diagnosis of disk degeneration from MR images is largely manual, which is time-consuming and labor-intensive, resulting in high cost. Recently, CNNs were developed for spine image analysis [20, 21] with the hope that machine learning algorithms may eventually serve as an AI doctor to automatically make a diagnosis.

We conducted this robustness study to investigate the feasibility of using CNNs for automated spine disk degeneration assessment from images without human intervention. The key to the medical assessment is to obtain the shapes of the individual disks, from which medical features can be calculated to determine the level of disk degeneration [22]. To reconstruct the shape of a disk, either image segmentation or shape regression can be applied using a CNN. Disk image segmentation is to classify the pixels of an image to one of the two classes: disk vs. the other region. Disk shape regression is to obtain the coordinates of the points on disk boundary (contour) from the image of a disk. To this end, we developed a U-net style CNN for both image segmentation and shape regression, and we investigated the robustness of this CNN with different training strategies.

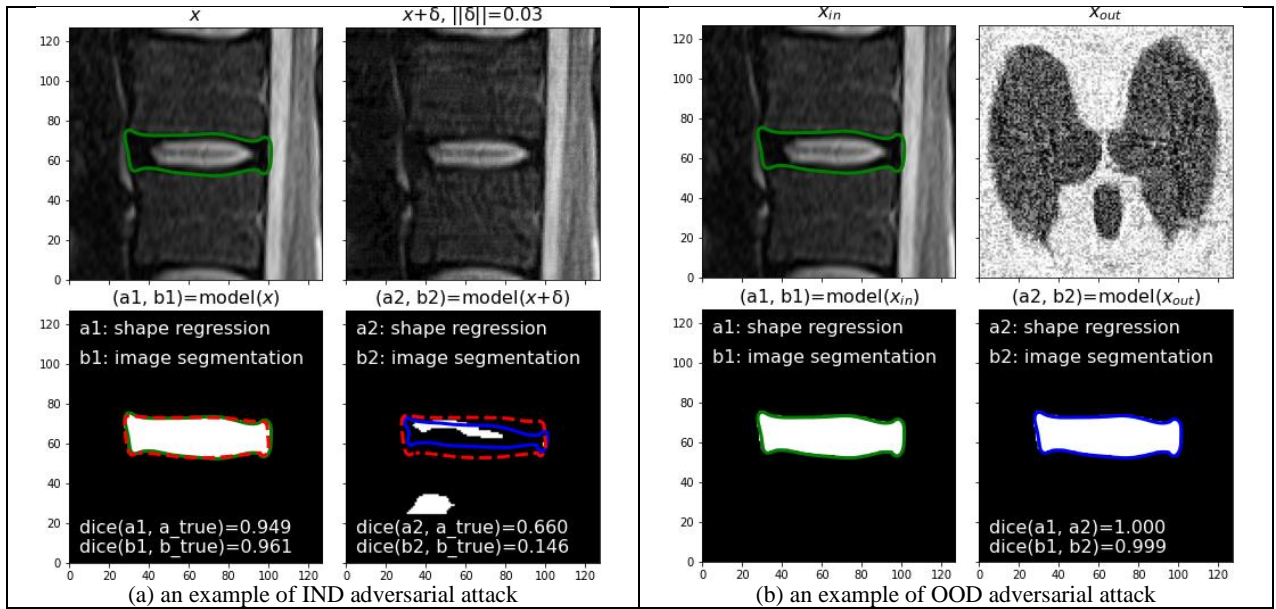

Figure. 1. Examples of IND and OOD adversarial attacks. The "ground-truth" disk boundary/contour is shown in red. Given a clean image as input, the boundary output of the CNN model is shown in green. Given an adversarial image as input, the boundary output of the model is shown in blue. A binary image is a segmentation output of the model.

Before we present our methods and results, we would like to make a distinction between IND adversarial attacks and OOD adversarial attacks, using the examples in Fig.1. Given a clean image $x$ of a disk as input, a CNN model will output the boundary/contour $y^{(a_1)}$ of the disk and the binary segmentation map $y^{(b_1)}$. A noisy image $x_\delta = x + \delta$ is obtained by adding a small amount of noise $\delta$ to $x$ through an IND adversarial attack. Given $x_\delta$ as input, the CNN model will output the boundary/contour $y^{(a_2)}$ of the disk and the binary segmentation map $y^{(b_2)}$. In Fig.1(a), the L-infinity (Linf) vector norm of the noise is 0.03, i.e., $\|\delta\| = 0.03$. The noise generated by an IND adversarial attack is small, but the CNN model changes its outputs significantly: the Dice similarity coefficient between the model-output segmentation and the ground-truth segmentation is reduced from 0.961 to 0.146, and the Dice similarity coefficient between the model-output contour and the ground-truth contour is reduced from 0.949 to 0.660. To compute Dice between two contours, the region enclosed by each contour is found first, and then the Dice between the two regions of the contours is calculated. For an IND adversarial attack, the goal is to generate an adversarial IND sample $x_\delta$ with a small amount of noise such that $x_\delta$ still looks like $x$ (by human eyes) but the CNN model will change its outputs significantly. For an OOD adversarial attack, the goal is to generate an adversarial OOD sample $x_{out}$ as input such that the corresponding outputs of the CNN model will be almost the same as the outputs of the model given the IND sample $x_{in} = x$ as input. In Fig.1(b), $x_{out}$ is a very noisy CT image of the lung, not looking like $x_{in}$, but the corresponding outputs are almost identical to those of the IND sample $x_{in}$. Since the goals of IND and OOD adversarial attacks are completely different, different strategies are needed to defend against the attacks. To handle IND adversarial attacks, a model should be insensitive to noise in the input. To handle OOD adversarial attacks, a model should be able to distinguish between IND and OOD samples.

## 2. METHODS

In this section, we provide the details of the CNN structure and dataset, explain the algorithm and loss functions for IND and OOD adversarial attacks, describe the process for IND adversarial training, and summarize the OOD deteciton method using reconstruction autoencoder.

### 2.1 Dataset

The dataset consists of de-identified lumbar spine MR images of 100 patients from University of Miami medical school. Three human experts manually annotated the boundaries and landmarks of the lumbar disks and vertebrae on the mid-sagittal MR image of each patient, by following the protocol in [22]. The best annotation (i.e., ground-truth) for each image was obtained through discussion to reach a consensus among the experts. Images and shapes from 80 patients were used for training; the images and shapes from the remaining patients were used for testing. Each lumbar image contains 5 disks. We cropped each image into squared regions, and each region contains a disk at the center. Then, each cropped region was resized to 128 ×128 pixels. A disk boundary is represented by a 2D contour that has 176 points.

We applied PCA to the shapes (i.e., contours) in the training set to build three statistical shape models (SSM) [23]: SSM-P3, SSM-P5, SSM-P10, which contain 3, 5, 10 principle components to cover 70.5%, 82.2%, and 91.4% of the total variations, respectively. Each SSM was used to generate an augmented training set of 640,000 virtual shapes. The augmented training set from SSM-P10 contains much larger shape-variations than the augmented training set from SSM-P3. For each virtual shape $\tilde{s}$, a virtual image $\tilde{x}$ is generated in the following steps: (1) randomly select an image $x$ in the original training set, which is accompanied with the ground-truth shape $s^*$, (2) compute the thin-plate-spline (TPS) transform from $\tilde{s}$ to $s^*$, and (3) apply the TPS transform to image $x$ and the warped image $\tilde{x}$ is the virtual image of the virtual shape $\tilde{s}$. In this way, we obtained three augmented training sets. An example is shown in Fig. 2.

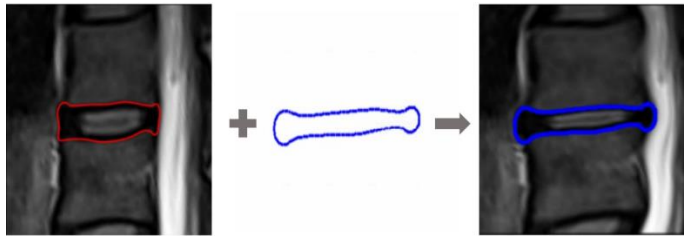

Figure 2. (left) a real image with the ground-truth disk shape. (middle) virtual shape. (right) virtual image.

### 2.2 CNN Structure

The structure of the CNN in the U-net style is shown in Fig. 3. The encoder is based on Resnet-18 [24], which contains convolutions layers and residual connections. The decoder has transposed-convolution blocks, and there are concatenative skip-connections from the encoder to the decoder. The input to the encoder is a disk image (128×128), and the output from the decoder is a binary segmentation map. The output of the encoder is connected to a block of fully-connected layers, which outputs the boundary/contour of the disk. LeakyReLU is used as activation. GroupNorm [25] is used for normalization as a replacement of BatchNorm [26] that often makes the network unstable for regression.

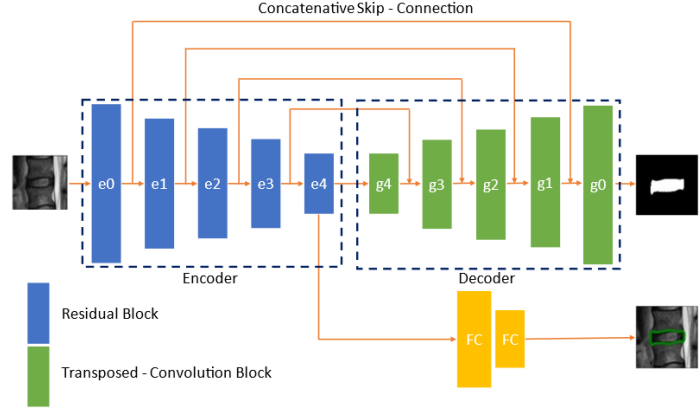

Figure 3. The structure of the CNN for shape (boundary) regression and image segmentation.

To train the CNN model for regression and segmentation, the loss function $L(x)$ is used:

$$L_{reg}(x) = \frac{1}{2 \times 176} \|\hat{s} - s\| \quad (1)$$

$$L_{seg}(x) = 0.5(1 - Dice(\hat{m}, m)) + 0.5 CE(\hat{m}, m) \quad (2)$$

$$L(x) = L_{reg}(x) + L_{seg}(x) \quad (3)$$

$L_{reg}(x)$ is the shape regression loss. $Dice$ is the function to compute dice similarity coefficient. $CE$ is the function to compute cross-entropy. $L_{seg}(x)$ is the image segmentation loss, consisting of Dice loss and CE loss. $s$ is the ground-truth contour (i.e., an array of the coordinates of the 176 points on the contour), and $\hat{s}$ is the model output for the input $x$. $m$ is the ground-truth segmentation map, and $\hat{m}$ is the model output for the input $x$.

### 2.3 The Algorithm for IND and OOD Adversarial Attacks

| Algorithm 1: Adversarial Attack (IND or OOD) |
|---|
| **Input**: |
| $x$, a sample (e.g., an image) in a dataset. |
| $f(x)$, the neural network for regression or segmentation/classification. |
| $J$, the objective function of the attack, which will be maximized. |
| $\varepsilon$, the maximum perturbation measured by Lp vector norm. |
| $N$, the total number of iterations. |
| $\alpha$, the learning rate of the optimizer (e.g., Adamax) |
| $x_{init}$, a sample for algorithm initialization |
| **Output**: an adversarial sample $x_\varepsilon$, s.t. $J(x_\varepsilon) \gg J(x)$ and $\|x_\varepsilon - x\|_p \leq \varepsilon$ |
| **Process:** |
| 1: generate a random noise $\xi$ with $\|\xi\|_p \leq \varepsilon$ |
| 2: initialize $x_\varepsilon = clip(x_{init} + \xi)$ |
| 3: **for** n from 1 to N: |
| 4: $\quad x_\varepsilon \leftarrow clip\left(x_\varepsilon + \alpha \cdot h(J'(x_\varepsilon))\right)$, where $J'(x) = \frac{\partial J}{\partial x}$ |
| Note: |
| The $clip$ operation ensures that $\|x_\varepsilon - x\|_p \leq \varepsilon$. The $clip$ operation also ensures that pixel values stay within the feasible range (e.g. 0 to 1). |
| If L-inf vector norm is used, $h(J')$ is the sign function; and if L2 vector norm is used, $h(J')$ is a function that normalizes $J'$ by its L2 norm. |

The Algorithm for adversarial attacks is described above. Although IND and OOD adversarial attacks have completely different goals, the same algorithm with different objective functions can be used to implement the attacks. The Algorithm is based on projected-gradient descent (PGD) [27], which is widely used for robustness evaluation [28] against IND adversarial attacks. We recently applied the Algorithm for OOD adversarial attacks that are largely underexplored [17].

The goal of the Algorithm is to generate an adversarial sample such that the objective function for the adversarial attack is maximized. For an IND adversarial attack, the objective function should measure the difference between an output of the model and the corresponding ground-truth, and in our application, the Algorithm will try to enlarge the error of shape output from the CNN model by adding a small amount of noise $\delta$ to the clean image $x$, as shown in Fig.1(a). The magnitude of the noise $\delta$ is constrained (i.e., $\|\delta\|_p \leq \varepsilon$) so that the noisy image, $x_\varepsilon = \delta + x$, looks like the clean image $x$. We set $x_{init} = x$. To evaluate the robustness of a CNN model, two objectives, $J_{IND\_reg}$ and $J_{IND\_seg}$ are used in the Algorithm for shape regression and image segmentation, respectively:

$$J_{IND\_reg} = \|\hat{s} - s\|_2^2 \tag{4}$$

$$J_{IND\_seg} = 1 - Dice(\hat{m}, m) \tag{5}$$

where $s$ is the ground-truth contour and $\hat{s}$ is the model output for the input $x_\varepsilon$; $m$ is the ground-truth segmentation and $\hat{m}$ is the model output for the input $x_\varepsilon$.

For an OOD adversarial attack, as shown in Fig. 1(b), we set $x_{init}$ to be an OOD sample not similar to any sample in the dataset. For example, $x_{init}$ could be a CT image, and $x$ is a lumbar disk image. The Algorithm will modify $x_{init}$ to obtain an OOD adversarial sample $x_{out} = x_\varepsilon$ such that the model outputs for the input $x_\varepsilon$ will be almost the same as the model outputs for the input $x$. Thus, to evaluate the ability of a CNN model or a CNN-based method to detect OOD samples, the two objectives, $J_{OOD\_reg}$ and $J_{OOD\_seg}$, for OOD adversarial attacks to regression and segmentation are chosen to be:

$$J_{OOD\_reg} = -\|\hat{s} - s\|_1 \tag{6}$$

$$J_{OOD\_seg} = Dice(\hat{m}, m) \tag{7}$$

### 2.4 IND Adversarial Training to Defend Against IND Adversarial Attacks

A natural idea to improve robustness against noise is to add noise to the images and train the CNN model on clean and noisy images. Therefore, to defend against IND adversarial attacks, we could add IND adversarial samples to the training set, and train the model on clean and noisy/adversarial samples, which is known as adversarial training and effective for many image classification applications [28, 29]. It is very time-consuming to generate adversarial samples using the Algorithm, and therefore we set $N$, the total number of iterations to be 20. The noise level $\varepsilon$ is set to 0.07 (Linf norm). $\alpha$ is set to 0.01. The loss function for adversarial training is:

$$L_{adv\_rs} = 0.5L(x) + 0.5\left(0.5L(x_\varepsilon^{(reg)}) + 0.5L(x_\varepsilon^{(seg)})\right) \tag{8}$$

where $L$ is defined by Eq.(3). $x_\varepsilon^{(reg)}$ is an adversarial sample generated by only attacking the regression output, and $x_\varepsilon^{(seg)}$ is another adversarial sample generated by only attacking the segmentation output.

We also applied two variants of the adversarial training, and the loss functions are:

$$L_{adv\_r} = 0.5L(x) + 0.5L(x_\varepsilon^{(reg)}) \tag{9}$$

$$L_{adv\_s} = 0.5L(x) + 0.5L(x_\varepsilon^{(seg)}) \tag{10}$$

Using the loss function $L_{adv\_r}$, the adversarial training aims to improve shape regression robustness. Using the loss function $L_{adv\_s}$, the adversarial training aims to improve image segmentation robustness.

As a weak form of adversarial training, we performed model training with uniform random noise, i.e., adding uniform random noise to the training images ($x_{rand} = x + noise$). We set the maximum noise amplitude to 0.07 (Linf form). The loss function for model training is:

$$L_{rand} = 0.5L(x) + 0.5L(x_{rand}) \tag{11}$$

### 2.5 Reconstruction-based OOD Detection to Defend Against OOD Adversarial Attacks

For medical image analysis, reconstruction autoencoders have been used for OOD detection with the assumption that the reconstruction error of an OOD sample will be much higher than the reconstruction errors of IND samples [18]. In our application, the CNN can be used for reconstruction-based OOD detection after a simple modification: adding another output channel in addition to the segmentation output, and this channel will output the reconstructed version $x_{rec}$ of the input $x$. If the reconstruction error $\|x_{out\_rec} - x_{out}\|$ of an ODD sample $x_{out}$ is larger than the reconstruction errors (i.e., $\|x_{in\_rec} - x_{in}\|$) of many samples in the training set, then $x_{out}$ is identified to be OOD. OOD detection is classification, and reconstruction error is classification score. By varying the classification threshold in a large range, the area under the receiver operating characteristic curve (AUROC) can be obtained to measure the performance of OOD detection [30].

AUROC is in the range of 0 to 1: if AUROC is less than 0.5, then OOD detection is ineffective and no better than a random guess. For model training, a reconstruction loss $L_{rec}$ needs to be added (e.g., added to $L$ in Eq.(3)):

$$L_{rec}(x) = \frac{1}{128 \times 128} \sum_{i=1}^{128 \times 128} |x_{rec}[i] - x[i]| \quad (12)$$

Here, $x_{rec}[i]$ is the value of the pixel-$i$ of the reconstruction output from the model. $x[i]$ is the value of the corresponding pixel of the input image.

## 3. RESULTS

### 3.1 Robustness Against IND Adversarial Attacks

To study model robustness against IND adversarial attacks, we trained 15 models that have the same structure (Fig. 3). For clarity, we gave these models different names. The name prefix of a model could be "P3", "P5" or "P10", which means the data generated from SSM-P3, SSM-P5, or SSM-P10 were used for training the model. "P3_std", "P5_std", and "P10_std" are the models trained with the loss in Eq.(3) only on clean data. "P3_rand", "P5_rand", and "P10_rand" are the models trained with the loss $L_{rand}$. "P3_adv_rs", "P5_adv_rs", and "P10_adv_rs" are the models trained with the loss $L_{adv\_rs}$ (i.e., adversarial training). "P3_adv_r", "P5_adv_r", and "P10_adv_r" are the models trained with the loss $L_{adv\_r}$. "P3_adv_s", "P5_adv_s", and "P10_adv_s" are the models trained with the loss $L_{adv\_s}$. The Adamax optimizer was used with the default parameters. The number of training epochs is 100, and batch size is 64.

For model evaluation, the noise level $\varepsilon$ is varied in a large range (0.01, 0.03, 0.05, 0.07, 0.1, 0.2), $\alpha$ is set to $\varepsilon/5$, and the maximum number of iterations $N$ is set to 100 in the Algorithm. Given a noise level, the shape regression performance of a model is measured by $Error\_reg$ and $DICE\_reg$, defined by

$$Error\_reg = \frac{1}{K_{test} i_{max}} \sum_{k=1}^{K_{test}} \sum_{i=1}^{i_{max}} dist(\hat{s}_k[i], s_k[i]) \quad (13)$$

$$DICE\_reg = \frac{1}{K_{test}} \sum_{k=1}^{K_{test}} Dice(\hat{s}_k, s_k) \quad (14)$$

$K_{test}$ is the number of the samples in the test set. $i_{max}$ is the number of points on a disk boundary, and $dist(\hat{s}_k[i], s_k[i])$ is the Euclidean distance between the point-$i$ on the $k$-th ground-truth shape $s_k$ and the corresponding point of the shape $\hat{s}_k$ output from the model. The image segmentation performance of a model is measured by $DICE\_seg$, defined by

$$DICE\_seg = \frac{1}{K_{test}} \sum_{k=1}^{K_{test}} Dice(\hat{m}_k, m_k) \quad (15)$$

The results are reported in Table 1 to 3 and visualized in Fig. 4 to 7. Examples are shown in Fig. 8&9.

Table 1: Shape regression errors ($Error\_reg$) of the models under different adversarial noise levels.

| noise | 0 | 0.01 | 0.03 | 0.05 | 0.07 | 0.1 | 0.2 |
|---|---|---|---|---|---|---|---|
| P03_std | 2.4913 | 4.0939 | 8.5369 | 10.1408 | 10.4531 | 10.525 | 10.5399 |
| p05_std | 2.0531 | 3.6349 | 10.024 | 15.2504 | 17.3851 | 18.6632 | 19.7239 |
| P10_std | 1.6886 | 3.917 | 12.781 | 21.5228 | 23.0385 | 23.527 | 24.3022 |
| P03_rand | 2.4726 | 3.8805 | 8.5904 | 10.6784 | 11.3548 | 11.652 | 12.6683 |
| p05_rand | 2.0398 | 3.4223 | 10.5357 | 18.1591 | 20.3922 | 21.5689 | 22.2194 |
| P10_rand | 1.684 | 3.6932 | 12.4964 | 22.6029 | 24.4443 | 25.2174 | 25.9216 |
| P03_adv_r | 2.6603 | 2.7958 | 3.1141 | 3.5179 | 4.0906 | 5.1555 | 8.4548 |
| P05_adv_r | 2.3019 | 2.5395 | 3.0886 | 3.7282 | 4.464 | 5.9477 | 9.3167 |
| P10_adv_r | 2.0098 | 2.3218 | 2.9138 | 3.5319 | 4.281 | 5.9074 | 9.5416 |
| P03_adv_s | 2.6959 | 2.9068 | 3.4149 | 4.0121 | 4.7687 | 6.1286 | 9.6913 |
| P05_adv_s | 2.1874 | 2.5595 | 3.4835 | 4.5297 | 5.7151 | 7.5313 | 11.3804 |
| P10_adv_s | 1.9462 | 2.5385 | 3.8151 | 5.046 | 6.2884 | 8.2526 | 18.1757 |
| P03_adv_rs | 2.657 | 2.8121 | 3.1788 | 3.6429 | 4.2769 | 5.4411 | 8.7731 |
| P05_adv_rs | 2.1857 | 2.4476 | 3.0946 | 3.8609 | 4.7454 | 6.2276 | 9.7687 |
| P10_adv_rs | 1.9608 | 2.3597 | 3.1337 | 3.9293 | 4.7708 | 6.2132 | 9.6021 |

Table 2: Shape regression accuracy ($DICE_{reg}$) of the models under different adversarial noise levels.

| noise | 0 | 0.01 | 0.03 | 0.05 | 0.07 | 0.1 | 0.2 |
|---|---|---|---|---|---|---|---|
| P03_std | 0.8772 | 0.8487 | 0.7063 | 0.6429 | 0.6235 | 0.6256 | 0.6216 |
| p05_std | 0.9064 | 0.8651 | 0.593 | 0.2895 | 0.1853 | 0.1377 | 0.1106 |
| P10_std | 0.9315 | 0.8906 | 0.5445 | 0.1888 | 0.1366 | 0.1224 | 0.105 |
| P03_rand | 0.8776 | 0.8522 | 0.7169 | 0.6266 | 0.5832 | 0.5638 | 0.5024 |
| p05_rand | 0.9063 | 0.8699 | 0.5943 | 0.2065 | 0.1282 | 0.1023 | 0.0917 |
| P10_rand | 0.9319 | 0.8959 | 0.5834 | 0.1897 | 0.1375 | 0.1154 | 0.0965 |
| P03_adv_r | 0.8705 | 0.8663 | 0.8573 | 0.8455 | 0.8287 | 0.7991 | 0.6977 |
| P05_adv_r | 0.8986 | 0.8885 | 0.8656 | 0.8403 | 0.8093 | 0.7407 | 0.6248 |
| P10_adv_r | 0.9186 | 0.9076 | 0.8857 | 0.8602 | 0.8311 | 0.7518 | 0.5721 |
| P03_adv_s | 0.8733 | 0.8685 | 0.8594 | 0.85 | 0.8348 | 0.8016 | 0.6727 |
| P05_adv_s | 0.906 | 0.8954 | 0.8698 | 0.8422 | 0.8124 | 0.7523 | 0.5795 |
| P10_adv_s | 0.9247 | 0.9142 | 0.8912 | 0.8643 | 0.8307 | 0.7504 | 0.3013 |
| P03_adv_rs | 0.8714 | 0.8671 | 0.8586 | 0.846 | 0.8273 | 0.7951 | 0.6916 |
| P05_adv_rs | 0.9059 | 0.8965 | 0.8708 | 0.8419 | 0.8064 | 0.74 | 0.6057 |
| P10_adv_rs | 0.9224 | 0.9109 | 0.8856 | 0.8579 | 0.8256 | 0.765 | 0.6131 |

Table 3: Image segmentation accuracy ($DICE_{seg}$) of the models under different adversarial noise levels.

| noise | 0 | 0.01 | 0.03 | 0.05 | 0.07 | 0.1 | 0.2 |
|---|---|---|---|---|---|---|---|
| P03_std | 0.8822 | 0.7846 | 0.177 | 0.0268 | 0 | 0 | 0 |
| p05_std | 0.9161 | 0.8328 | 0.297 | 0.0537 | 0.0108 | 0 | 0 |
| P10_std | 0.9446 | 0.847 | 0.2907 | 0.0468 | 0.0013 | 0 | 0 |
| P03_rand | 0.8812 | 0.7951 | 0.2027 | 0.0343 | 0 | 0 | 0 |
| p05_rand | 0.914 | 0.8444 | 0.3888 | 0.0795 | 0.0206 | 0 | 0 |
| P10_rand | 0.9441 | 0.8579 | 0.3196 | 0.0498 | 0.007 | 0 | 0 |
| P03_adv_r | 0.8863 | 0.8598 | 0.7866 | 0.6532 | 0.4601 | 0.129 | 0 |
| P05_adv_r | 0.9009 | 0.8247 | 0.6385 | 0.4216 | 0.1869 | 0.0396 | 0 |
| P10_adv_r | 0.9228 | 0.8245 | 0.5922 | 0.3451 | 0.1838 | 0.0569 | 0 |
| P03_adv_s | 0.8755 | 0.8698 | 0.8574 | 0.8416 | 0.8142 | 0.7264 | 0.1144 |
| P05_adv_s | 0.9119 | 0.9001 | 0.872 | 0.8367 | 0.7929 | 0.6611 | 0.137 |
| P10_adv_s | 0.9321 | 0.9152 | 0.8827 | 0.8473 | 0.8007 | 0.6607 | 0.1936 |
| P03_adv_rs | 0.8742 | 0.8685 | 0.856 | 0.8387 | 0.8069 | 0.7133 | 0.1178 |
| P05_adv_rs | 0.9106 | 0.8958 | 0.8657 | 0.8263 | 0.7751 | 0.614 | 0.0701 |
| P10_adv_rs | 0.9266 | 0.9097 | 0.8755 | 0.833 | 0.7827 | 0.6378 | 0.0538 |

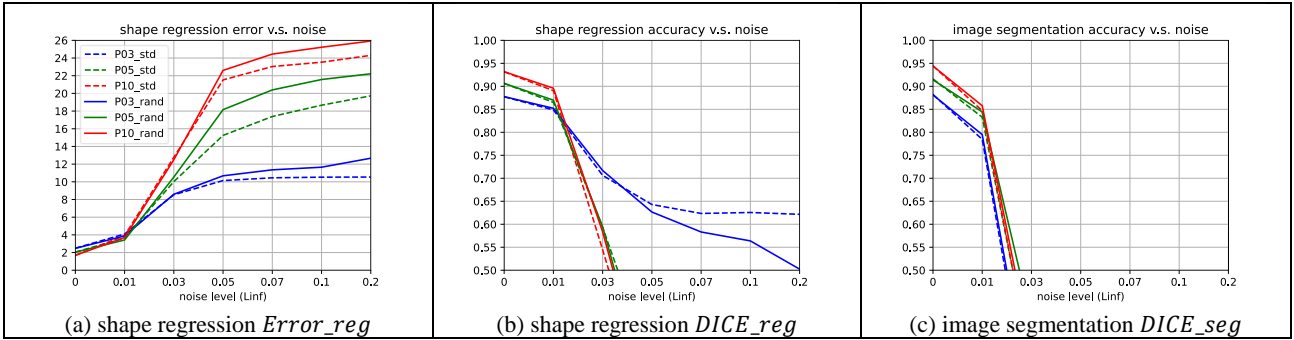

Figure 4. The performance of the models trained only on clean data dropped quickly (DICE < 0.5) even when the noise was very small (0.03). The robustness did not improve by training on noisy images with uniform random noises.

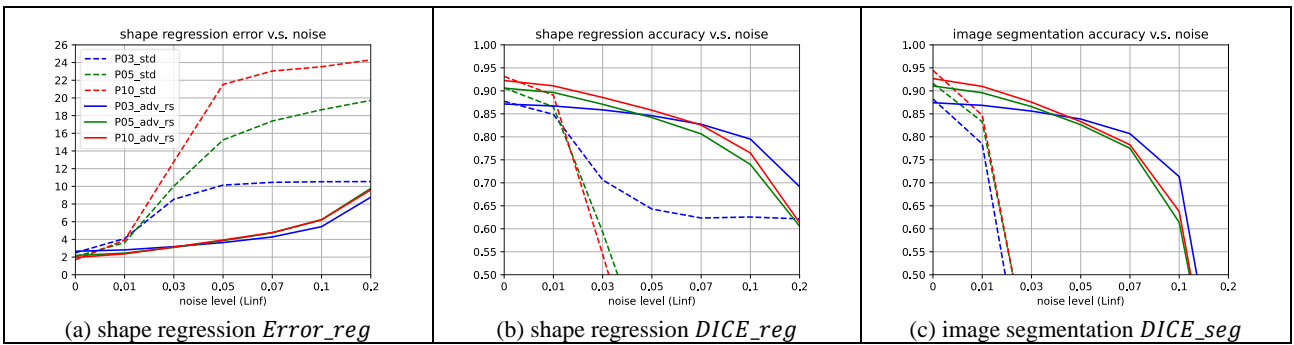

Figure 5. The shape regression and image segmentation robustness were significantly improved by IND adversarial training with the loss $L_{adv\_rs}$. When noise is small, the model (P10_adv_rs) trained on a larger dataset performed better than the model (P3_adv_rs) trained on a smaller dataset. By comparing P10_adv_rs and P10_std, it can be seen that the adversarial training had a side effect that model performance on clean data (noise level is 0) may drop.

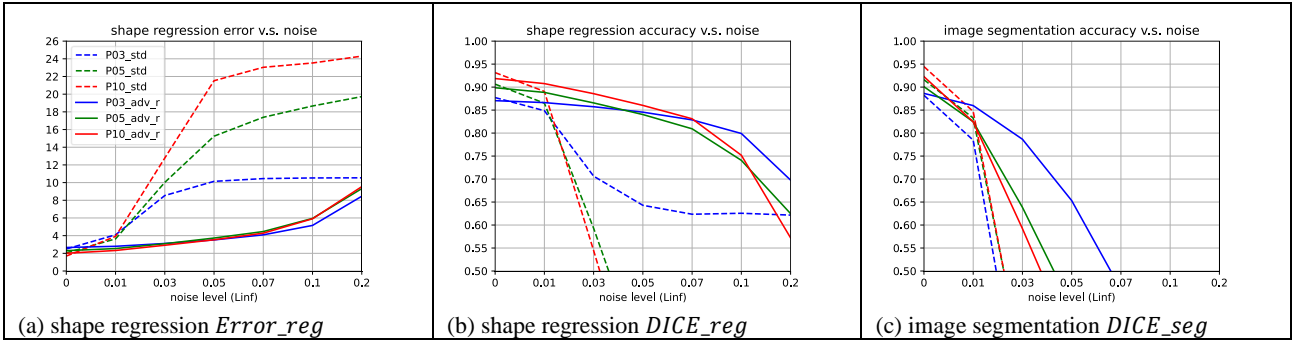

Figure 6. The shape regression robustness was significantly improved by IND adversarial training with the loss $L_{adv\_r}$, and the image segmentation robustness was improved only slightly by the adversarial training. When noise is small, the model (P10_adv_r) trained on a larger dataset performed better than the model (P3_adv_r) trained on a smaller dataset. By comparing P10_adv_r and P10_std, it can be seen that the adversarial training had a side effect that model performance on clean data (noise level is 0) dropped.

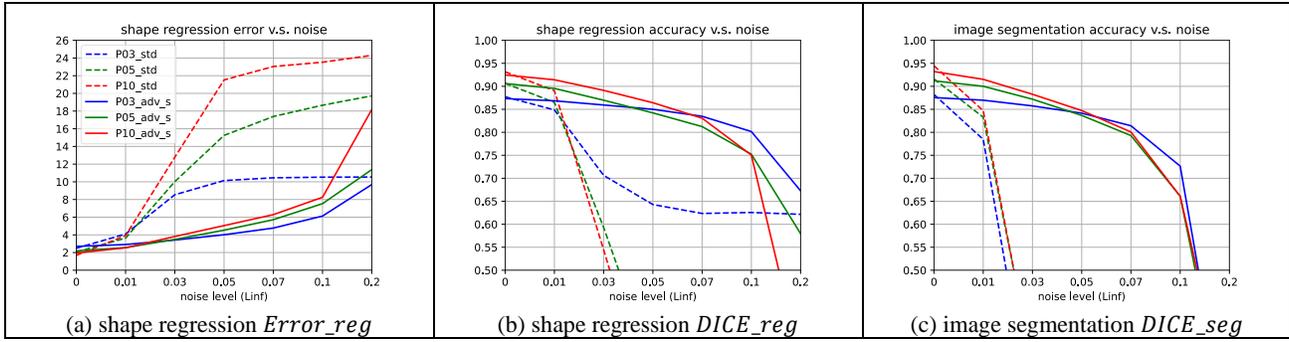

(a) shape regression $Error_{reg}$  (b) shape regression $DICE_{reg}$  (c) image segmentation $DICE_{seg}$

Figure 7. The image segmentation robustness was significantly improved by IND adversarial training with the loss $L_{adv\_s}$. The shape regression robustness was also significantly improved by the adversarial training. When noise is small, the model (P10_adv_s) trained on a larger dataset performed better than the model (P3_adv_s) trained on a smaller dataset. By comparing P10_adv_s and P10_std, it can be seen that the adversarial training led to a decrease of model performance on clean data.

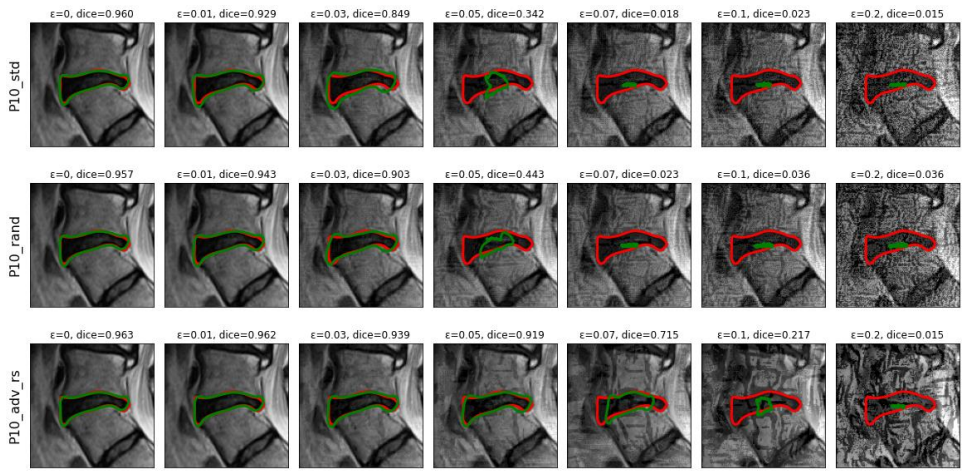

Figure 8. Shape regression outputs of three models under the IND adversarial attack with the loss $J_{IND\_reg}$. The ground-truth shape is shown in red color. Each output shape contour is shown in green color.

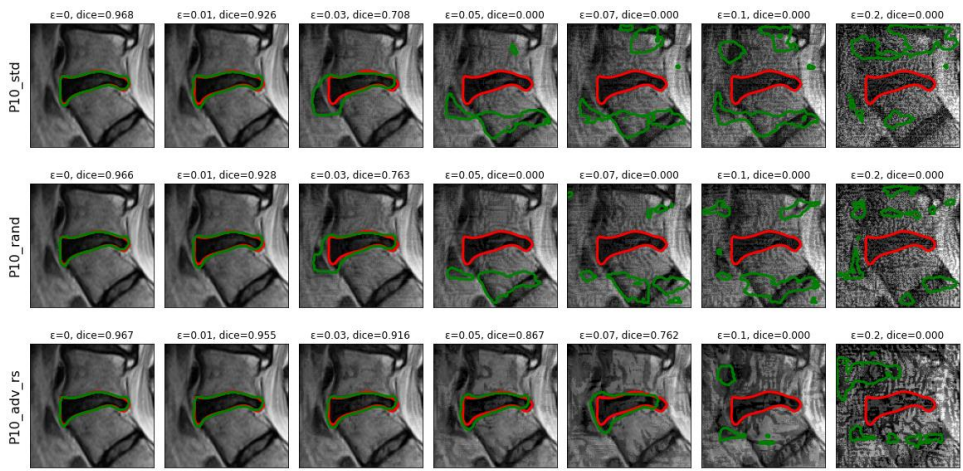

Figure 9. Image segmentation outputs of three models under the IND adversarial attack with the loss $J_{IND\_seg}$. The boundaries of the regions in the output segmentation map are traced and visualized in green color. The ground-truth shape is shown in red color.

## 3.2 Robustness Against OOD Adversarial Attacks

We first studied how the CNN model trained with the standard loss (i.e., P10_std) would behave under OOD adversarial attacks. In the Algorithm, we set $x_{init}$ to be a CT image of the lung, which is taken from an open CT dataset [31]. An example of the OOD adversarial attack to the CNN model has been shown in Fig. 1(b). During the OOD adversarial attack with $J_{OOD\_reg}$ or $J_{OOD\_seg}$, for each sample $x$ in the test set, a corresponding adversarial sample $x_{out}$ was generated, and the model output for $x$ and the model output for $x_{out}$ were compared using DICE, and two histograms of the DICE scores were obtained and shown in Fig. 10, which indicate that the CNN model was unable to "see" the difference between the IND samples in the test set and the OOD samples (noisy CT images generated by the OOD adversarial attack).

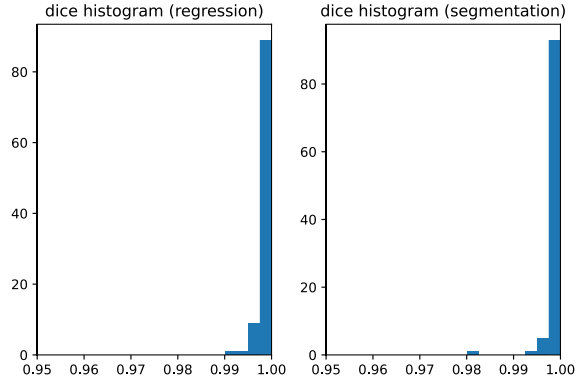

Figure 10. The results of OOD adversarial attack to the CNN model on the test set.

We then evaluated the reconstruction-based OOD detection method (Section 2.5). The data from SSM-P10 were used for model training with different loss functions. The histograms of reconstruction errors are shown in Fig. 11, which indicate that the method is unable to distinguish between IND and OOD samples, and its AUROC is less than 0.5.

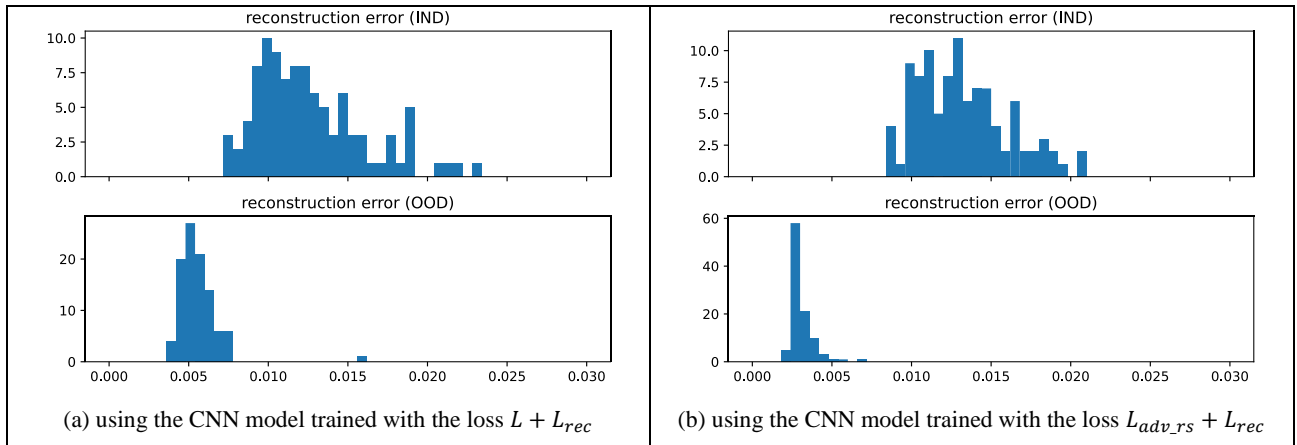

(a) using the CNN model trained with the loss $L + L_{rec}$    (b) using the CNN model trained with the loss $L_{adv\_rs} + L_{rec}$

Figure 11. Histograms of reconstruction errors show that the reconstruction-based OOD detection method is ineffective.

By examining the examples from the OOD adversarial attacks, as shown in Fig. 12, we observed a very interesting phenomenon: each adversarial sample from the model trained with the loss $L + L_{rec}$ is equal to the original clean OOD sample (a CT image of the lung) plus some random noises, whereas each adversarial sample from the model trained with the loss $L_{adv\_rs} + L_{rec}$ is the combination of the original clean OOD sample, random noises, and an "object" whose shape is similar to a disk shape. This observation suggests that the model trained by IND adversarial training may have learned robust image-features of disk shapes, compared to the model trained with the standard loss.

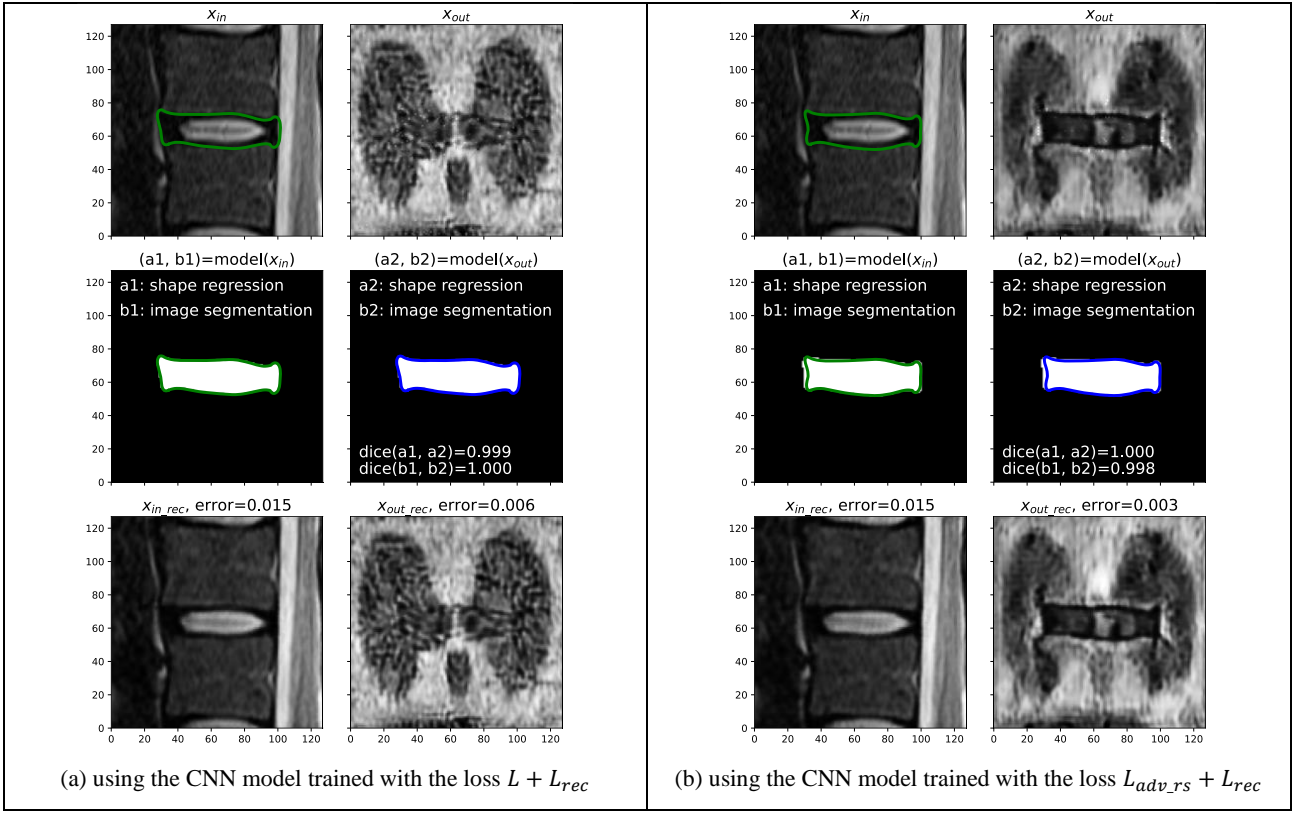

Figure 12. Examples from the OOD adversarial attacks to the two models trained with two different loss functions.

## 4. DISCUSSION

The models trained with the standard loss (P3_std, P5_std, and P10_std) were vulnerable to IND adversarial attacks, and their performance for shape regression and image segmentation dropped significantly on noisy data (i.e., IND adversarial samples) even when the noise is imperceptible (Fig.1(a), Fig.8, and Fig. 9). Training the models with random noises (P3_rand, P5_rand, and P10_rand) did not improve robustness to IND adversarial attacks (Fig. 4). IND adversarial training significantly improved robustness of the models (P3_adv_rs, P5_adv_rs, and P10_adv_rs) for regression and segmentation (Fig. 5). By focusing on the segmentation output with the loss $L_{adv\_s}$, not only segmentation robustness but also regression robustness was significantly improved (Fig. 7); however, by focusing on the regression output with the loss $L_{adv\_r}$, only regression robustness was significantly improved (Fig. 6). Under IND adversarial training, the models trained on a "larger" training set (from SSM-P10) had better robustness than those trained on a "smaller" training set (from SSM-P3) when the noise level was small.

OOD adversarial attacks are underexplored and have a significantly different goal, compared to IND adversarial attacks. The results (Fig.1(b) and Fig. 10) show that the model trained with the standard loss (P10_std) cannot detect any difference in image patterns between the disk images (IND samples) and the noisy CT images (OOD samples). The reconstruction-based OOD detection method can be easily fooled by the OOD adversarial attack, as shown in Fig. 11. Combined with our recent work on OOD adversarial attacks [17], we can make a safe conclusion that the current OOD detection methods, which have been evaluated by us, cannot defend against OOD adversarial attacks.

Although IND adversarial training can be very effective, OOD adversarial training is computationally infeasible because the space of OOD samples is significantly larger than the space of IND samples. An OOD sample could be an X-ray image of the chest, a gray-scale image of a human face, or even a random noise image, as long as it does not look like the samples in the training set. The boundary between IND and OOD adversarial attacks could be blurry when the noise level is high. As shown in Fig. 8&9, when the noise level ε is higher than 0.1, the noisy images look very different from the real images and should be treated as OOD samples, which is the reason that we set ε to 0.07 for IND adversarial training.

## 5. CONCLUSION

We conducted the robustness study of the U-net style CNN for lumbar spine disk shape reconstruction from MR images, and the results imply that the current neural network technologies are not robust enough for automated spine disk analysis from images without human intervention. Although IND adversarial training can be used to improve robustness against small input noises from IND adversarial attacks, it still remains to be a challenge to defend against OOD adversarial attacks, which means AI doctor on a human-expert level still remains to be a distant dream. Therefore, current neural network-based medical systems should serve only as assistants to human doctors. Given the fact that humans can easily detect the OOD samples, there must be some fundamental differences between the human brain and current artificial neural networks, which is worth further investigation.

## SOURCE CODE

The source code of this study is publicly available at https://github.com/jiasongchen/SPIE-2021

## ACKNOWLEDGEMENT

This work was supported in part by Amazon AWS Machine Learning Research Award.